# Stimulated wave of polarization in spin chains

G.B. Furman [1], S.D. Goren [1], J.-S. Lee [2], A.K. Khitrin [2], V.M. Meerovich [1], and V.L. Sokolovsky [1]

[1] Department of Physics, Ben Gurion University, Beer Sheva 84105, Israel

[2] Department of Chemistry, Kent State University, Kent, Ohio 44242-0001, USA

Stimulated wave of polarization, triggered by a flip of a single spin, presents a simple model of quantum amplification. Previously, it has been found that such wave can be excited in a 1D Ising chain with nearest-neighbor interactions, irradiated by a weak resonant transverse field. Here we explore models with more realistic Hamiltonians, in particular, with natural dipole-dipole interactions. Results of simulations for 1D spin chains and rings with up to nine spins are presented.

## 1. Introduction

Recently, it has been demonstrated that, in a one-dimensional Ising chain with nearest-neighbor interactions, irradiated by a weak resonant transverse field, a wave of flipped spins can be triggered by a single spin flip[1]. This analytically solvable model illustrates a principle of quantum amplification and can play an important role in design of quantum devices for amplifying signals from a single quantum object[2-7]. The Hamiltonian of the model[1] is

$$H = \frac{\omega_0}{2}\sum_{i=1}^{N}\sigma_i^z + \omega_1\sum_{i=1}^{N}\sigma_i^x \cos\omega_0 t + \frac{J}{4}\sum_{i=1}^{N-1}\sigma_i^z\sigma_{i+1}^z, \qquad (1)$$

where $\omega_0$ is the energy difference ($\hbar = 1$) between the excited and ground states of an isolated spin (qubit), $J$ is the interaction constant, $\omega_1 \ll J \ll \omega_0$ is the amplitude of irradiation, $\sigma^z$ and $\sigma^x$ are the Pauli operators. The idea of operation of this model is the following. When a spin is at the either end of the chain or when it has two neighbors in the same state, interaction with the neighbor(s) makes the spin off-resonant and the irradiation field does not change its state. When the two neighbors of a spin are in different states, the resonant irradiation field flips the spin. Therefore, if all spins are in the same state, the state of the entire system is stationary. If the first spin is flipped, its neighbor becomes resonant and flips, then the next spin flips, and so on, resulting in "quantum domino" dynamics.

At $\omega_1 \ll J \ll \omega_0$ the Hamiltonian (1) can be replaced by the effective (secular) Hamiltonian

$$\tilde{H}_{\text{secular}} = \frac{\omega_1}{4}\sum_{i=2}^{N-1}\sigma_i^x\left(1 - \sigma_{i-1}^z\sigma_{i+1}^z\right), \qquad (2)$$

which is the time-independent part of the Hamiltonian in the interaction frame ($\omega_1 \sum_{i=1}^{N} \sigma_i^x \cos\omega_0 t$ is treated as a perturbation). The terms of the three-spin effective Hamiltonian of Eq. (2) have a simple interpretation: the spin-flipping operator $\sigma_i^x$ for the spin $i$ is "turned off" when its two neighbors are in the same state and $\left(1 - \left\langle \sigma_{i-1}^z \sigma_{i+1}^z \right\rangle\right)$ is zero. The dynamics with the Hamiltonian (2) can be solved exactly[1] for the two initial states: when all the spins are up (it is an eigenstate of the Hamiltonian (2) and, therefore, does not change), and when the first spin of the chain is flipped.

The idealized Hamiltonian (1) is difficult to realize in experimental systems. More realistic Hamiltonians may include interactions beyond the nearest neighbors and may not be limited to ZZ interactions. In this work we study how these factors affect generation of a stimulated polarization wave and demonstrate that the "quantum domino" dynamics can be implemented with realistic Hamiltonians, including the natural dipole-dipole interactions.

## 2. Long-range interactions

Let us consider a linear chain or ring of $N$ spins coupled by long-range dipolar interactions. First, we will discuss the weak coupling limit[8], when the difference in precession frequencies of interacting spins exceeds the interaction strength and the interaction is truncated to ZZ-terms only. It will be also assumed that all spins are simultaneously irradiated at their resonance frequencies. This slightly artificial model will be used to clarify the role of long-range interactions. Later we will consider a more realistic model with two sorts of spins under two-frequency irradiation. The truncated dipole-dipole Hamiltonian is

$$H_{dd} = \sum_{m>n}^{N} D_{mn} I_m^z I_n^z, \qquad (3)$$

where $D_{mn} = \dfrac{D_1}{(m-n)^3}$ for a linear chain, $D_{mn} = D_1 \left[ \dfrac{\sin(\frac{\pi}{N})}{\sin(\frac{\pi}{N}(m-n))} \right]^3$ for a ring, and $D_1$ is the coupling strength between the nearest spins. When only interactions with $M$ nearest neighbors are taken into account the Hamiltonian in the rotating frame takes the form

$$H = \omega_1 \sum_{m=1}^{N} I_m^x + \sum_{q=1}^{M} D_q \sum_{m=1}^{N-q} I_m^z I_{m+q}^z, \qquad (4)$$

where for the linear chain $D_q = D_{m,m+q} = \dfrac{D_1}{q^3}$ and for the ring $D_q = D_{m,m+q} = D_1 \left[ \dfrac{\sin(\frac{\pi}{N})}{\sin(\frac{\pi q}{N})} \right]^3$. In the limit $\omega_1 \ll D_{mn}$ one can calculate the effective Hamiltonian for the chain

$$H_{chain}^{eff} = \frac{\omega_1}{2} \sum_{k=M+1}^{N-M} I_k^x \Pi_{q=1}^{M} \left(1 - 4 I_{k+q}^z I_{k-q}^z \right) + H_{end}^{eff}, \qquad (5)$$

where $H_{end}^{eff}$ is the part of the effective Hamiltonian for the first $M$ and the last $M$ spins at the ends of the chain. We present the explicit forms of $H_{end}^{eff}$ only for the cases $M = 1$ and $M = 2$. At $M = 1$

$$H_{end}^{eff} = D_1 \left( I_1^z + I_N^z \right) + \frac{\omega_1}{4} I_1^x \left( 1 - 2 I_2^z \right) + \frac{\omega_1}{4} I_N^x \left( 1 - I_{N-1}^z \right), \qquad (6)$$

and at $M = 2$

$$H_{end}^{eff} = \frac{7D_1}{16}(I_1^z + I_N^z) - \frac{D_1}{16}(I_2^z + I_{N-1}^z) + \frac{\omega_1}{4}I_1^x(1 - 2I_2^z I_3^z + I_2^z - I_3^z)$$
$$+ \frac{\omega_1}{8}I_2^x(1 - 4I_1^z I_3^z)(1 + 2I_4^z) + \frac{\omega_1}{4}I_N^x(1 - 2I_{N-2}^z I_{N-1}^z + I_{N-3}^z - I_{N-2}^z) \quad (7)$$
$$+ \frac{\omega_1}{8}I_{N-1}^x(1 - 4I_{N-2}^z I_N^z)(1 + 2I_{N-3}^z)$$

For the ring the effective Hamiltonian is

$$H_{ring}^{eff} = \frac{\omega_1}{4}\sum_{k=1}^{N} I_k^x \Pi_{q=1}^{M}(1 - I_{k+q}^z I_{k-q}^z), \quad (8)$$

where the cyclic conditions $I_0^z = I_N^z$, $I_{N+p}^z = I_p^z$ are assumed.

We used numerical simulations to calculate time dependencies of the individual spin polarizations $P_k(t) = Tr\{I_k^z e^{-itH}\rho(0)e^{itH}\}$ and the total spin polarization $P(t) = \sum_{k=1}^{N} P_k(t)$, where $\rho(0)$ is the initial density matrix. Assuming that the spin system is initially in the ground state, except for the first spin flipped, the initial density matrix can be written as

$$\rho(0) = \otimes_{k=1}^{N}\rho_k(0), \quad (9)$$

where $\rho_1(0) = \begin{vmatrix} 0 & 0 \\ 0 & 1 \end{vmatrix}_1$ and $\rho_k(0) = \begin{vmatrix} 1 & 0 \\ 0 & 0 \end{vmatrix}_k$.

The results for the seven-spin chain with the nearest-neighbor interactions ($M = 1$) are shown in Fig. 1a. Spin dynamics is very close to the one obtained with the effective Hamiltonian (2) and shown in Fig. 1b.

At $M > 1$ the effective Hamiltonians (5) without the last term does not describe propagation of the stimulated polarization wave, initiated by a flip of the first spin. The reason is that interaction with next neighbors spoils the resonance condition and, in the limit $\omega_1 \to 0$, the transverse field cannot flip the spins. However, at intermediate values of $\omega_1$ when the amplitude of the transverse field is smaller than interaction between the nearest neighbors but larger than interaction between next-nearest neighbors, the off-resonance effect is suppressed and the polarization wave can be launched. The results for $\omega_1 = 0.15\, D_1$ and $M = 2$ are shown in Fig. 2a. Inclusion of interactions with farther spins does not cause qualitative changes in the dynamics but makes the wave somewhat less pronounced. The results for all spin coupled ($M = N - 1$) are shown in Fig. 2b.

## 3. Full dipole-dipole interaction

In a strong external field $\omega_0 = \gamma H_0$, the secular part of the dipole-dipole interaction

$$H_{dd} = \sum_{m>n} D_{mn}\left[ I_m^z I_n^z - \frac{1}{2}\left(I_m^x I_n^x + I_m^y I_n^y\right)\right], \qquad (10)$$

with the coupling constants

$$D_{mn} = \frac{\gamma^2}{2 r_{mn}^3}\left(1 - 3\cos^2\theta_{mn}\right), \qquad (11)$$

where $r_{mn}$ is the distance between spins $m$ and $n$, $\theta_{mn}$ is the angle between the vector $\vec{r}_{mn}$ and the external magnetic field $\vec{H}_0$, includes interactions between all three spin components. XX- and YY-interactions cause mutual flips of the spins even without the transverse resonant field ($\omega_1 = 0$). The spin-diffusion wave, propagating from the first flipped spin, is shown in Fig. 3a. However, this wave does not provide any amplification since the total Z-component of magnetization is conserved. With the dipolar interactions (10), stimulated wave, initiated by a flip of the first spin, cannot be launched at any value of $\omega_1$. The reason is the presence of strong X and Y local fields created by the nearest neighbors. These fields do not allow a resonance selection, which would depend on Z-components of the neighbor spins.

In an alternating spin chain ABAB…, when the difference of the resonance frequencies of spins A and B exceeds the strength of dipolar interaction between them (?), XX- and YY-interactions between nearest neighbors can be suppressed. As a result, interaction between the nearest neighbors will be reduced to ZZ-terms, while the interactions with farther spins can be suppressed by choosing an intermediate value of $\omega_1$, similar to the cases presented in Fig. 2. It is supposed that both types of the spins are simultaneously irradiated at their resonance frequencies with the fields of the same amplitude $\omega_1$. It is interesting to note that some spin-dynamic problems can be solved exactly for alternating rings[9], chains[10], and even three-periodic chains[11]. Since at present the experimental tests are easier to perform with nuclear magnetic resonance (NMR), we made a simulation for the alternating chain of $^1$H and $^{19}$F nuclear spins with gyromagnetic ratios $\gamma_{^1H} = 42.58$ MHz/T and $\gamma_{^{19}F} = 40.08$ MHz/T, respectively. The results are presented in Fig. 3b. One can see that a good stimulated polarization wave is excited at $\omega_1 = 0.15\, D_1$.

**4. Spin rings**

The major difference between rings and 1D chains is that in a ring the first flipped spin has two neighbors and it initiates two stimulated waves propagating symmetrically in both directions. The results of simulation for a nine-spin ring are presented in Fig. 4. Fig. 4a shows the dynamics of polarizations of individual spins with the Hamiltonian (4) at $\omega_1 = 0.15\, D_1$. The results for the effective Hamiltonian (8) at $M = 1$ are displayed in Fig. 4b. Compared to linear chains, when the first flipped spin is at one of the ends (Figs. 1 and 2), each of the two stimulated waves has decreased amplitude. In addition, a polarization of the initially flipped spin is not "frozen" but increases with time. However, both of the models in Fig.4 demonstrate dynamics with efficient amplification, similar to that in linear spin chains.

## 5. Discussion

The most important feature of the spin dynamics described above is signal amplification, when a state of polarization of a single spin is converted into a total polarization of the spin system. Change in the total polarization (magnetization) of a spin system quantifies the efficiency of this system as a quantum amplifier. The results for different models studied in this work are summarized in Fig. 5. Figs. 5a and 5b show time dependences of the total polarization in seven-spin linear chains and in seven-spin rings, respectively. In all simulations, exclude simulation presented on Fig. 3a, the amplitude of the transverse resonant field was fixed at the value $\omega_1 = 0.15\ D_1$. Simulations showed that for the nearest-neighbor interactions, both linear and ring models demonstrated efficient amplification. When the long-range interactions are included, the performance depends on the amplitude of the resonant field $\omega_1$. With a proper choice of $\omega_1$, which should be smaller than the nearest-neighbor interactions but larger than the next-nearest interactions, it is possible to organize efficient amplification dynamics in 1D chains with long-range interactions. It is interesting that alternating chains of two spin species can work well even with full dipole-dipole couplings between spins. For small seven-spin rings, long-range couplings spoil the dynamics.

In conclusion, we have studied propagation of the polarization waves, triggered by a flip of a single spin, in one-dimensional spin chains and rings. It has been demonstrated that efficient amplification dynamics can be organized for natural dipole-dipole couplings when alternating chains of two spin species are used.

**Acknowledgments**

This work was supported by the US-Israel Binational Science Foundation.

**Figure captions**

Fig. 1. (a) Time dependences of the individual spin polarizations in a linear chain of seven spins calculated by using the Hamiltonian (4) with the nearest-neighbor interactions ($M = 1$) and $\omega_I = 0.15$ $D_1$; (b) time dependences of the individual

polarizations in a linear chain of seven spins calculated by using the zero-order average Hamiltonian (2).

Fig. 2. Time dependences of the individual spin polarizations in a linear chain of seven spins calculated by using the Hamiltonian (4) at $\omega_1 = 0.15\ D_1$ for (a) $M = 2$ and (b) $M = N - 1$.

Fig. 3. (a) Time dependences of the individual spin polarizations in a linear chain of seven dipolar-coupled spins calculated by using the Hamiltonian (10); (b) alternating chain of $^1$H and $^{19}$F nuclear spins at $\omega_1 = 0.15\ D_1$.

Fig. 4. (a) Time dependences of the individual spin polarizations in a ring of nine spins calculated by using the Hamiltonian (4) at $\omega_1 = 0.15\ D_1$; (b) time dependences of the individual polarizations in a ring of nine spins calculated by using the effective Hamiltonian (8) at $M = 1$.

Fig. 5. (a) Time dependences of the total spin polarizations in seven-spin chains: with the effective Hamiltonian (2) (solid), and, at $\omega_1 = 0.15\ D_1$, for ZZ couplings with $M = 1$ (dash), $M = 2$ (dots), $M = N - 1$ (dash-dot), and for the alternating spin chain with dipole-dipole couplings (dash-dot-dot); (b) time dependences of the total spin polarizations in seven-spin rings: with the effective Hamiltonian (8) at $M = 1$ (solid), and, at $\omega_1 = 0.15\ D_1$, for ZZ couplings with $M = 1$ (dashed), $M = 2$ (dots), $M = N - 1$ (dash-dot), and for the alternating spin chain with dipole-dipole couplings (dash-dot-dot).

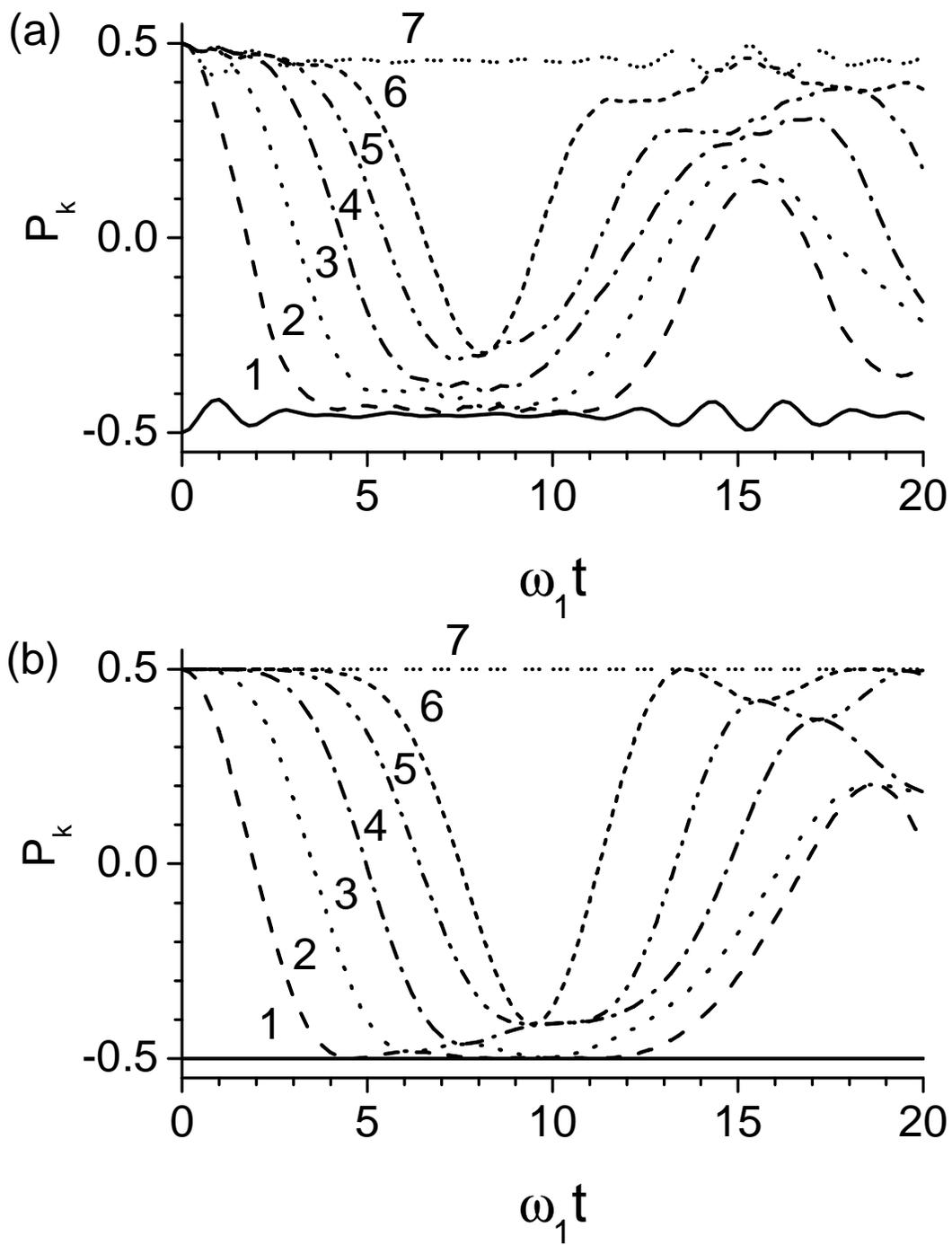

Fig. 1

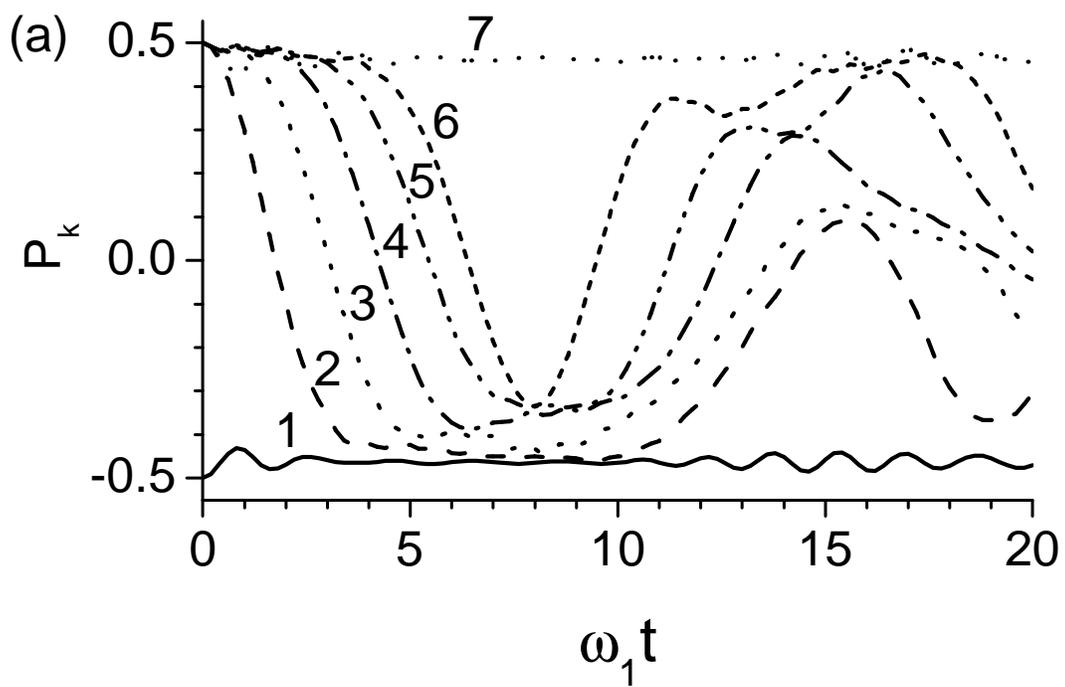

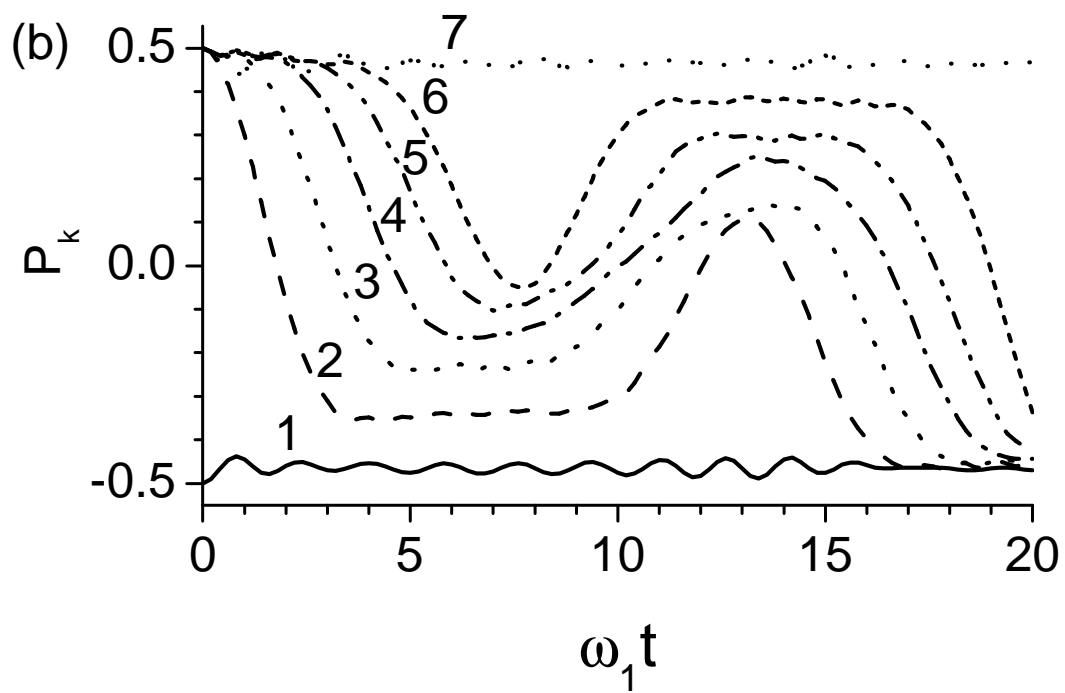

Fig. 2

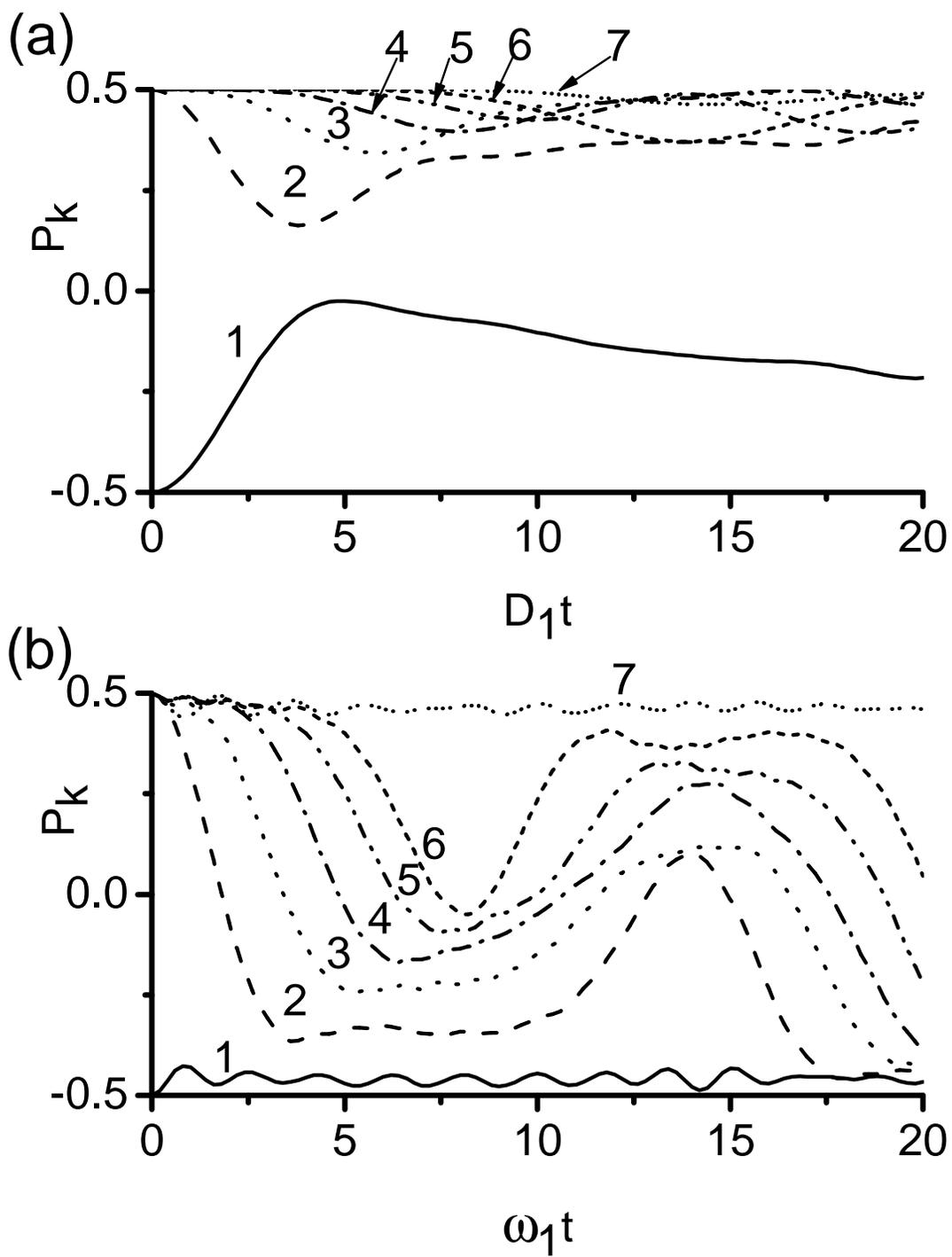

Fig. 3

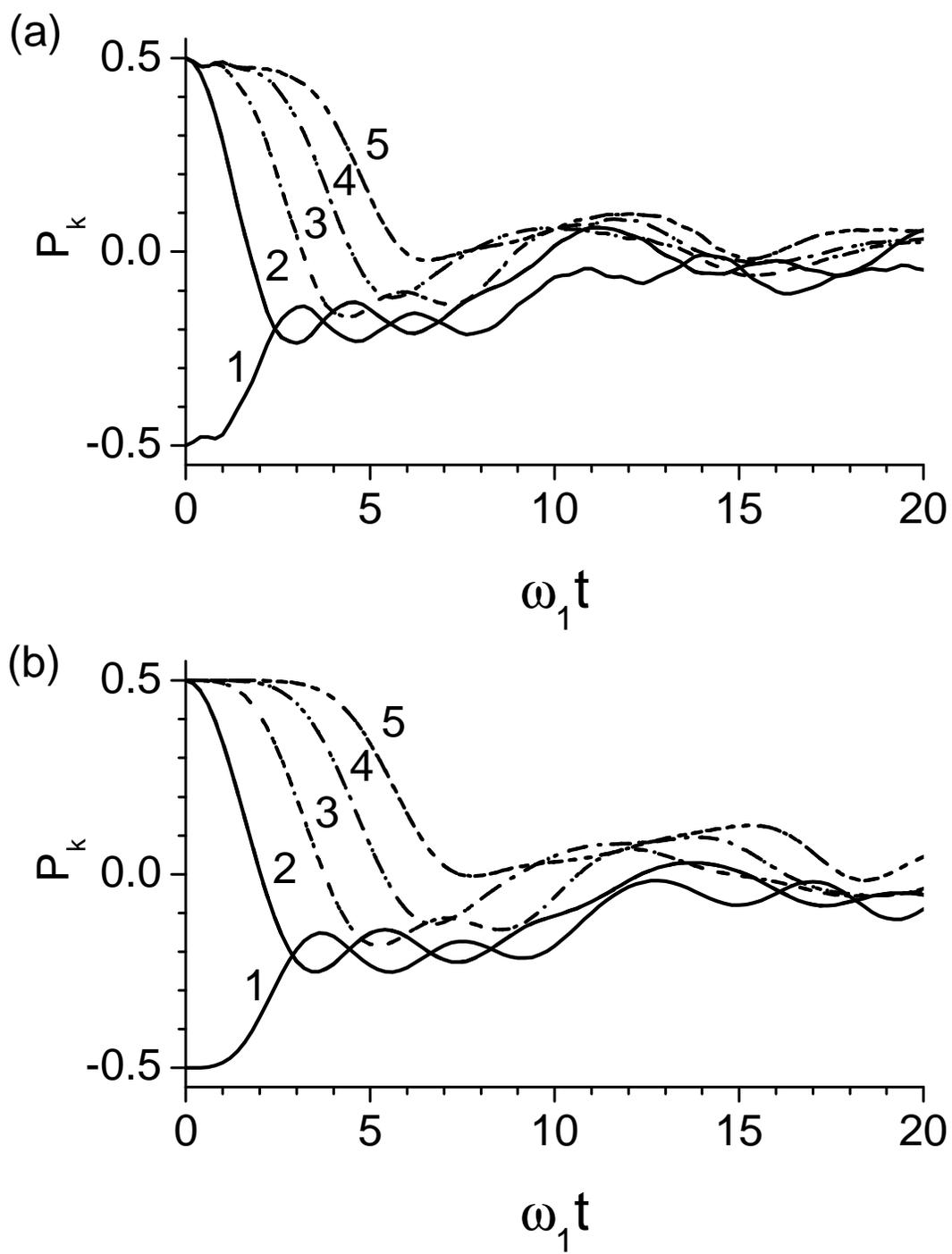

Fig. 4

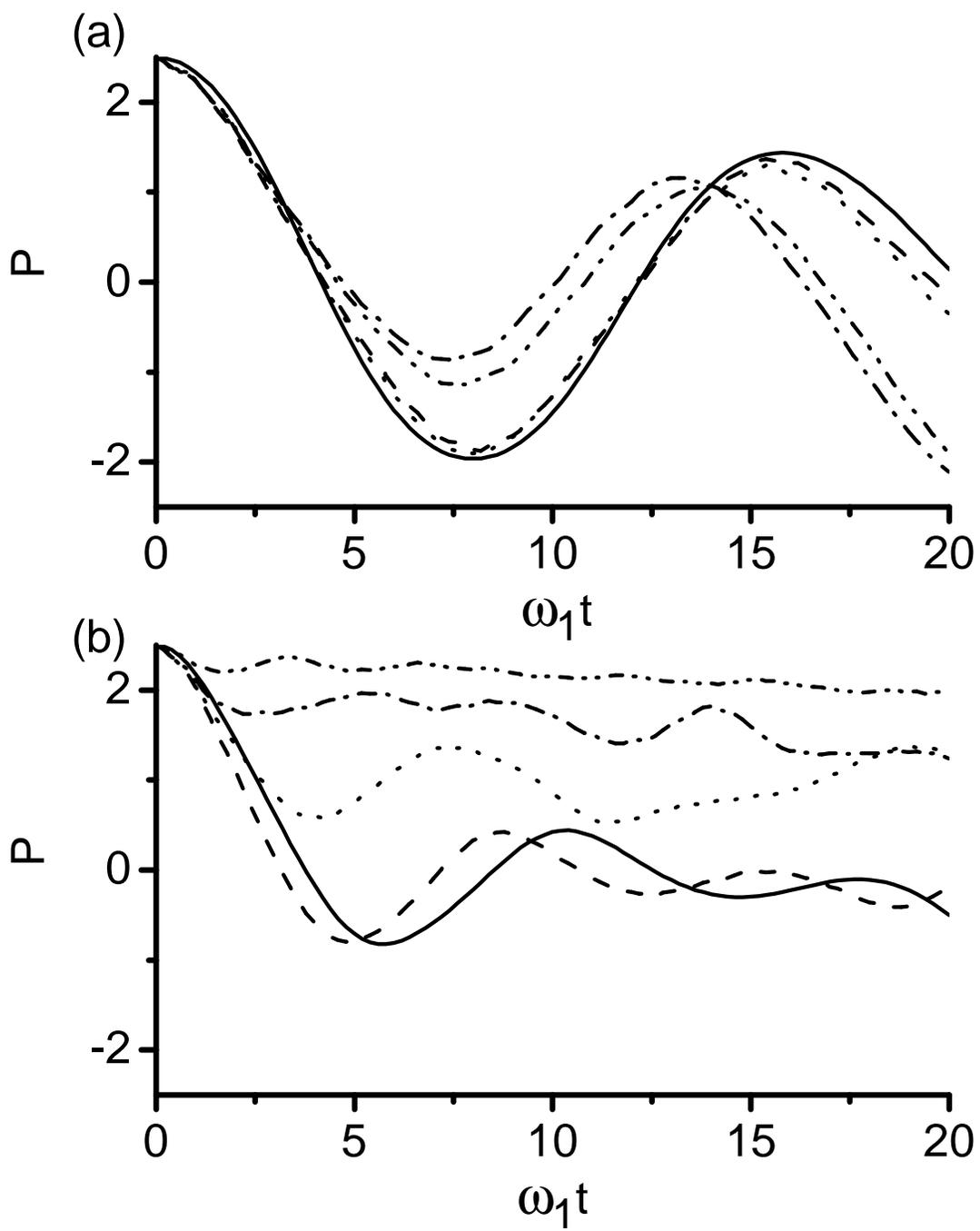

Fig. 5